\documentclass[twocolumn,amsmath,amssymb]{PoS}
\usepackage{braket}
\usepackage{amsmath}
\usepackage{amsfonts}
\usepackage{amssymb}
\usepackage{graphics}
\usepackage{graphicx}
\usepackage{capt-of}
\usepackage{multicol}
\usepackage{multirow}

\usepackage{csquotes}
\MakeOuterQuote{"}

\title{Recent measurements of branching fractions and $CP$ asymmetries of charmless hadronic $B$ meson decays at Belle}

\ShortTitle{Recent measurements of branching fractions and $CP$ asymmetries of charmless hadronic $B$ meson decays at Belle}

\author{\speaker{A. B. Kaliyar}%
        \thanks{for the Belle Collaboration.}\\
       IIT Madras\\
       E-mail: \email{basithkaliyar@physics.iitm.ac.in}}

\abstract{Hadronic $B$ decays without a charm quark constitute a powerful probe to search for physics beyond the standard model as well as to provide constraints on $CP$ violation parameters. We report final measurements from Belle of the branching fractions and $CP$ asymmetries for the decays $B^{0} \rightarrow \pi^{0} \pi^{0}$ and  $B^{\pm} \rightarrow K^{+} K^{-} \pi^{\pm}$. The $B^{0} \rightarrow \pi^{0} \pi^{0}$ measurements enable us to improve constraints  on the angle $\phi_{2}$ of the CKM unitarity triangle. For $B^{\pm} \rightarrow K^{+} K^{-} \pi^{\pm}$ we measure $CP$ asymmetry as a function of the invariant mass of the $K^{+}K^{-}$ system, where we find a strong evidence for large $CP$ violation and a large increase in yield at low mass region. This result challenges conventional theoretical approaches as it requires a large enhancement in both tree and loop level diagrams in the same small region of phase-space. }

\FullConference{The European Physical Society Conference on High Energy Physics\\
                 5-12 July\\
                 Venice, Italy}

\begin{document}

\section{Introduction}

Charmless decays of $B$ mesons to hadronic final states are suppressed in the standard model (SM), hence they constitute a powerful probe to search for new physics as well as to provide constraints on $CP$ violation parameters. Any significant deviation from SM expectations of the branching fraction or $CP$ asymmetry  will be a hint towards new physics. In this paper, we present measurements of branching fractions and $CP$ asymmetries for the charmless  hadronic $B$ decays $B^{\pm} \rightarrow K^{+} K^{-} \pi^{\pm}$ and  $B^{0} \rightarrow \pi^{0} \pi^{0}$ based on the full $\Upsilon(4S)$ data collected with the Belle detector at the KEKB asymmetric energy $e^{+} e^{-}$ collider [1]. 

\section{$B^{\pm} \rightarrow K^{+} K^{-} \pi^{\pm}$ decay}
 $B$ meson decays to three-body charmless hadronic final states of $ K^{+} K^{-} \pi^{+}$ [2] are dominated by the Cabbibo-suppressed $b\rightarrow u$ tree and $b\rightarrow d$ loop transitions. Large $CP$ asymmetries can occur in these decays, due to  interference between  SM tree and loop level diagrams with similar amplitudes. An unidentified structure has been measured by BaBar [3] and LHCb [4, 5] in the low $K^{+} K^{-}$ invariant mass spectrum of the $B^{+} \rightarrow K^{+} K^{-} \pi^{+}$ decay. Also, LHCb reported a non-zero inclusive $CP$ asymmetry of $-0.123 \pm 0.017 \pm 0.012 \pm 0.007$ (where, the first uncertainty is statistical, the second is the systematic, and the third is due to the $CP$ asymmetry  of the $B^{\pm} \rightarrow J/\psi^{} K^{\pm}$  reference mode) and a large unquantified local $CP$ asymmetry in the same mass region. These results suggest that final-state interactions can be a contributing factor to $CP$ violation [6, 7].  With this study, we attempt to quantify the $CP$ asymmetry and branching fraction as a function of the $K^{+} K^{-}$ invariant mass based on a data sample collected at the $\Upsilon(4S)$ resonance by the Belle detector comprising of $772 \times 10^{6}$ $B\bar{B}$,  which corresponds to an integrated luminosity of 711 fb$^{-1}$, and an additional 89.4 fb$^{-1}$ of off-resonance sample recorded at a center-of-mass energy around 60 MeV below the $\Upsilon (4S)$  resonance. 
 
%\begin{figure}
%\begin{center}
%\includegraphics[scale=0.4]{fig1a}
%\includegraphics[scale=0.4]{fig1b}\\
%\includegraphics[scale=0.4]{fig1c}
%\includegraphics[scale=0.4]{fig1d}
%\end{center}
%\caption{$B^{+} \rightarrow K^{+} K^{-} \pi^{+}$ Feynman diagrams (all Cabibbo-suppressed). (a) tree diagram, (b) W -exchange diagram leading to $KK^{*}$ states, (c) strong penguin diagram, and (d) electroweak penguin leading to the $\phi %\pi$ state}
%\end{figure}
%\subsection{Event Selection}
We combine two oppositely-charged kaons with a charged pion to reconstruct $B^{+} \rightarrow K^{+} K^{-} \pi^{+}$.
Charged kaons and pions  are identified based on a likelihood ratio obtained by combining information from the aerogel Cherenkov counters, time-of-flight counters,  and central drift chamber, $ L_{K/\pi} =\frac{L_{K}}{(L_{K} + L_{\pi})}$, where $L_{K}$ and $L_{\pi}$ are the likelihoods for the kaon and pion hypothesis, respectively. Tracks with $ L_{K/\pi} > 0.6$  are chosen as kaons and those with $ L_{K/\pi} < 0.4$ as pions.  Signal $B$ events are identified with two kinematic variables: the beam-energy constrained mass, $M_{bc}= \sqrt{E_{\rm{beam}}^{2}/c^{4} - |p_{B}/c|^{2}} $ and the energy difference, $\Delta E = E_{B} -  E_{\rm{beam}}$, where  $E_{\rm{beam}}$ is the beam energy and $E_{B}$ and $p_{B}$ are the energy and momentum of the $B$ candidate  in the center-of-mass frame, respectively. The fit region is defined as $M_{bc} > 5.24 ~\mathrm{GeV/}c^{2}$ and $-0.3 < \Delta E < 0.3 ~\rm{GeV}$, while the signal-enhanced region is defined as $5.27 <M_{bc} < 5.29  ~\mathrm{GeV/}c^{2}$ and $-0.05 < \Delta E < 0.05 ~\rm{GeV}$. When multiple $B$ candidates are present in an event, we choose the candidate with the best fit quality from the $B$ vertex fit. This is done for 19\% of events and the selection efficiency is 92\%.

%\subsection{Background study}
The dominant backgrounds are from $e^+ e^- \rightarrow q\bar{q}~(q= u, d, s, c)$ continuum process. The $B\bar{B}$ events are spherical in shape whereas the particles from continuum events are collimated into two back-to-back jets. We make use of  this difference in event topology by using a neural network [8] to combine several shape variables along with other properties of the event that distinguish $q\bar{q}$ from $B\bar{B}$ events. A requirement on the neural network output ($C_{NN}$ $>0.88$) is applied to suppress continuum background. This selection requirement is optimized by maximizing a figure of merit defined as $\frac{N_{S} }{ N_{S} + N_{B}}$, where $N_{S}$ ($N_{B}$) denotes the expected number of signal (background) events in the signal-enhanced region. Background contributions from $B$ decays mediated by the dominant $b \rightarrow c$ transition are investigated with an MC sample of such decays. To suppress these backgrounds, candidates for which the invariant mass of the $K^{+} K ^{-}$ or $K^{+} \pi^{-}$ system lies in the range  185\textendash188~MeV/$c^{2}$ are removed. This selection window corresponds to $ \pm 3.75 \sigma$ around the nominal $D^{0}$ mass [9], where $\sigma$ is the mass resolution. Backgrounds from charmless $B$ decays are studied with a large MC sample, where one of the $B$ mesons decays via a process with a known branching fraction. The study reveals  that a few modes contribute in the $M_{bc}$ signal region with a corresponding $\Delta E$ peak, denoted collectively as the "rare peaking" background. These peaking backgrounds are due to $K-\pi$ misidentification, which consist of $B^{+} \rightarrow K^{+} K^{-} K^{+}$, $B^{+} \rightarrow K^{+} \pi^{-} \pi^{+}$, and their intermediate resonant modes. Events that remain after removing the peaking components are called the "rare combinatorial" background.\\

%\subsection{Signal extraction}
The signal yield is extracted by performing a two-dimensional unbinned extended maximum likelihood fit in $M_{bc}$  and $\Delta E$ with the likelihood defined as
 \begin{equation}
   \mathcal{L} = \dfrac{e^{-\sum_{j} N_{j}}}{N!} \prod_{i} \Big[\sum_{j} N_{j} \mathcal{P}_{j}^{i} \Big], \ \ \mathrm{where} \ \ \ 
    \mathcal{P}_{j}^{i} = \dfrac{1}{2} (1-q_{i}.A_{CP})\times \mathcal{P}_{j} (M_{bc}^{i} , \Delta E^{i}),
 \end{equation}  
where $i$ denotes the event index, $N_{j}$  is the yield for the component $j$, $q_{i}$ is the charge of $B$ candidates ($q_{i}=\pm 1$ for $B^{\pm}$),  and $\mathcal{P}_{j}$ is the probability density function (PDF) corresponding to the component $j$.  Figure 1 shows the fit results of first two $M_{K^{+} K^{-}}$  bins in the signal-enhanced region. The resulting branching fraction and $CP$ asymmetry are [10]
\begin{equation}
\mathcal{B}(B^{+} \rightarrow K^{+} K^{-} \pi^{+} ) = (5.38 \pm 0.40 \pm 0.35) \times 10^{-6} 
\end{equation} 
and
\begin{equation}
A_{CP} = -0.182 \pm 0.071 \pm 0.016, 
\end{equation} 
where the quoted uncertainties are statistical and systematic, respectively.

To investigate the localized $CP$ asymmetry in the low $M_{K^{+} K^{-}}$ region, we determine the signal yield and $A_{CP}$ in bins of $M_{K^{+} K^{-}}$. The fitted results are shown in Table 1 and Fig. 2, where an excess of signal yield as well as a large $A_{CP}$ are seen in $M_{K^{+} K^{-}} < 1.5 ~\mathrm{GeV/}c^{2} $, confirming the observations by BaBar and LHCb. We find strong evidence for a large $CP$ asymmetry of $-0.90 \pm 0.17 \pm 0.03$ with a significance of $4.8\sigma$ for  $M_{K^{+} K^{-}} < 1.1 ~\mathrm{GeV/}c^{2}$.
\begin{figure}[]
\begin{small}
\begin{center}
\includegraphics[scale=0.37]{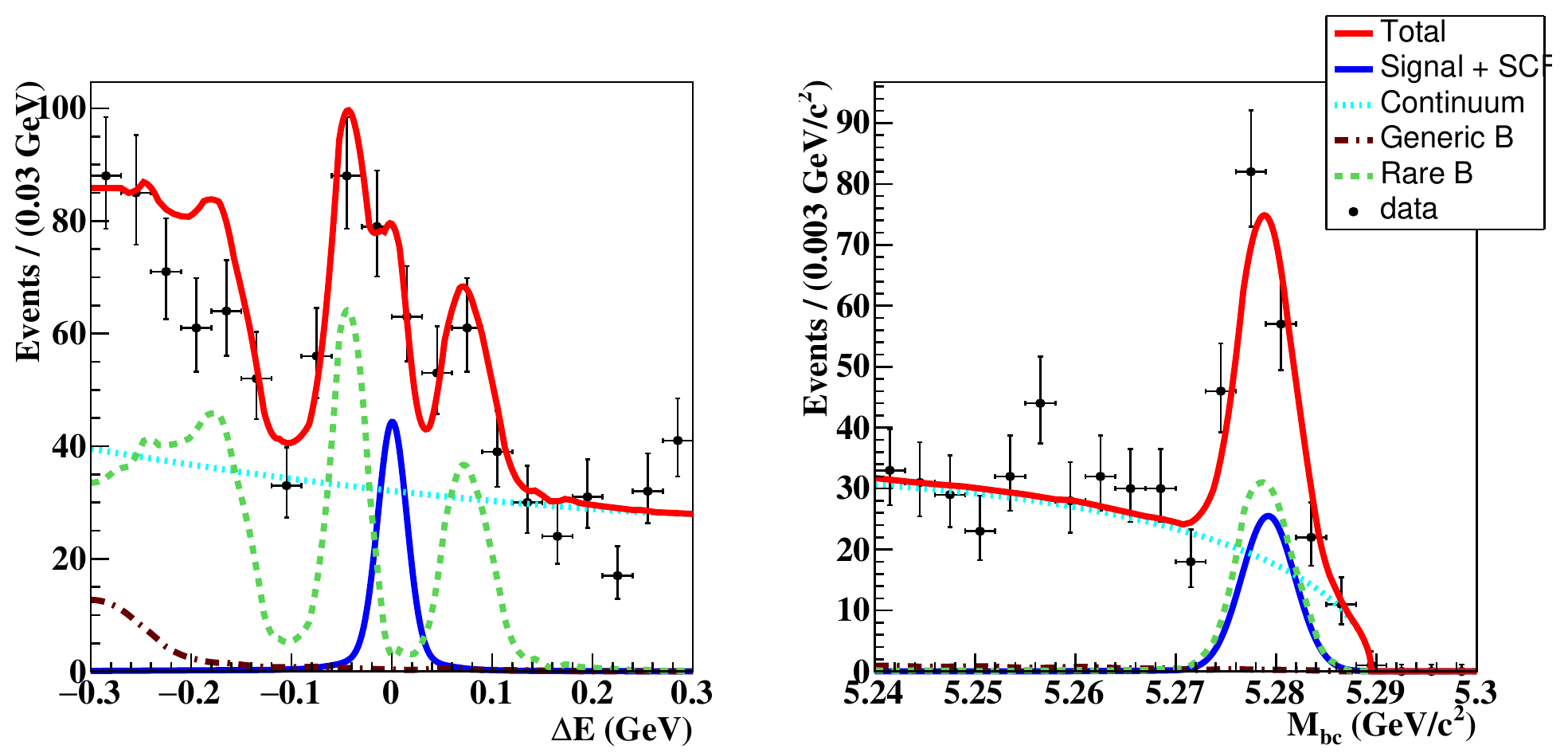}
\includegraphics[scale=0.37]{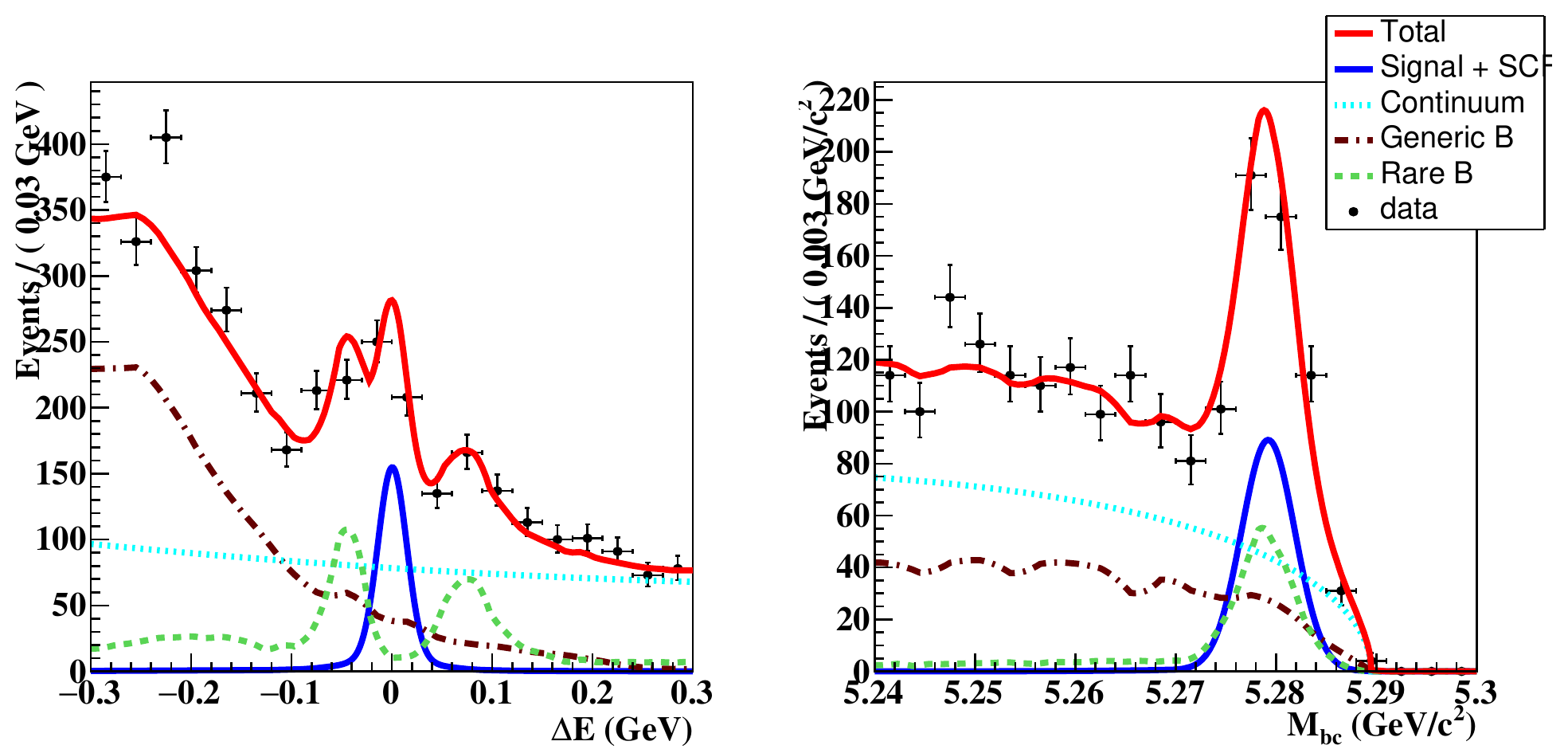}
\end{center}
\caption{Signal-enhanced projections of the $M_{bc}$-$\Delta E$ fit to data in the first (left) and second (right) $M_{K^{+} K^{-}}$  bins. Points with error bars are the data, red solid curves are the fit result, blue solid curves are the sum of the signal and the self cross-feed, cyan dotted curves are the continuum background, brown dash dotted curves are the generic $B$ backgrounds, and green dashed curves are the rare $B$ backgrounds.}
\end{small}
\end{figure}

\begin{table}
\caption{Signal yield, efficiency, differential branching fraction, and $\mathcal{A}_{CP}$ for individual $M_\mathrm{{KK}}$ bins \vspace*{0.05 in}}
\begin{center}
\hspace*{-0.25 in}
\begin{tabular}{  c  c  c  c  c }
\hline \hline  
\vspace{-0.09 in}
\\ $\textit{M}_\mathrm{{K^{+}K^{-}}} $ & $N_\mathrm{{sig}}$ & Eff. (\%) &$\textit{d} \mathrm{\mathcal{B}/\textit{dM} ~(\times 10^{-7}})$& ${A}_{CP}$ \\

$\mathrm{(GeV/c^{2})}$	&			&		&		& \\

\hline 
0.8\textendash1.1 &	~~$59.8\pm 11.4 \pm 2.6$ & 19.7 &$14.0\pm 2.7 \pm 0.8$ & $-0.90\pm 0.17 \pm 0.03$\\

1.1\textendash1.5 & $212.4\pm 21.3\pm 6.6$ &  19.3& $37.8\pm 3.8 \pm 1.9$ & $-0.16\pm 0.10 \pm 0.01$\\

1.5\textendash2.5 & ~$113.5\pm 26.7 \pm 18.0 $ & 15.6& $10.0\pm 2.3 \pm 1.6$ & $-0.15\pm 0.23 \pm 0.03$\\

2.5\textendash3.5 & $110.1 \pm 17.6 \pm 4.1 $ & 15.1& $10.0\pm 1.6 \pm 0.5$ & $-0.09\pm 0.16 \pm 0.01$\\

3.5\textendash5.3 & ~$172.6 \pm 25.7 \pm 6.87$  & 16.3& ~~$8.1\pm 1.2 \pm 0.5$ & $-0.05\pm 0.15 \pm 0.00$\\ 

\hline \hline 
\end{tabular}
\end{center}
\end{table}

\begin{figure}[]
\begin{small}
\begin{center}
\includegraphics[scale=0.25]{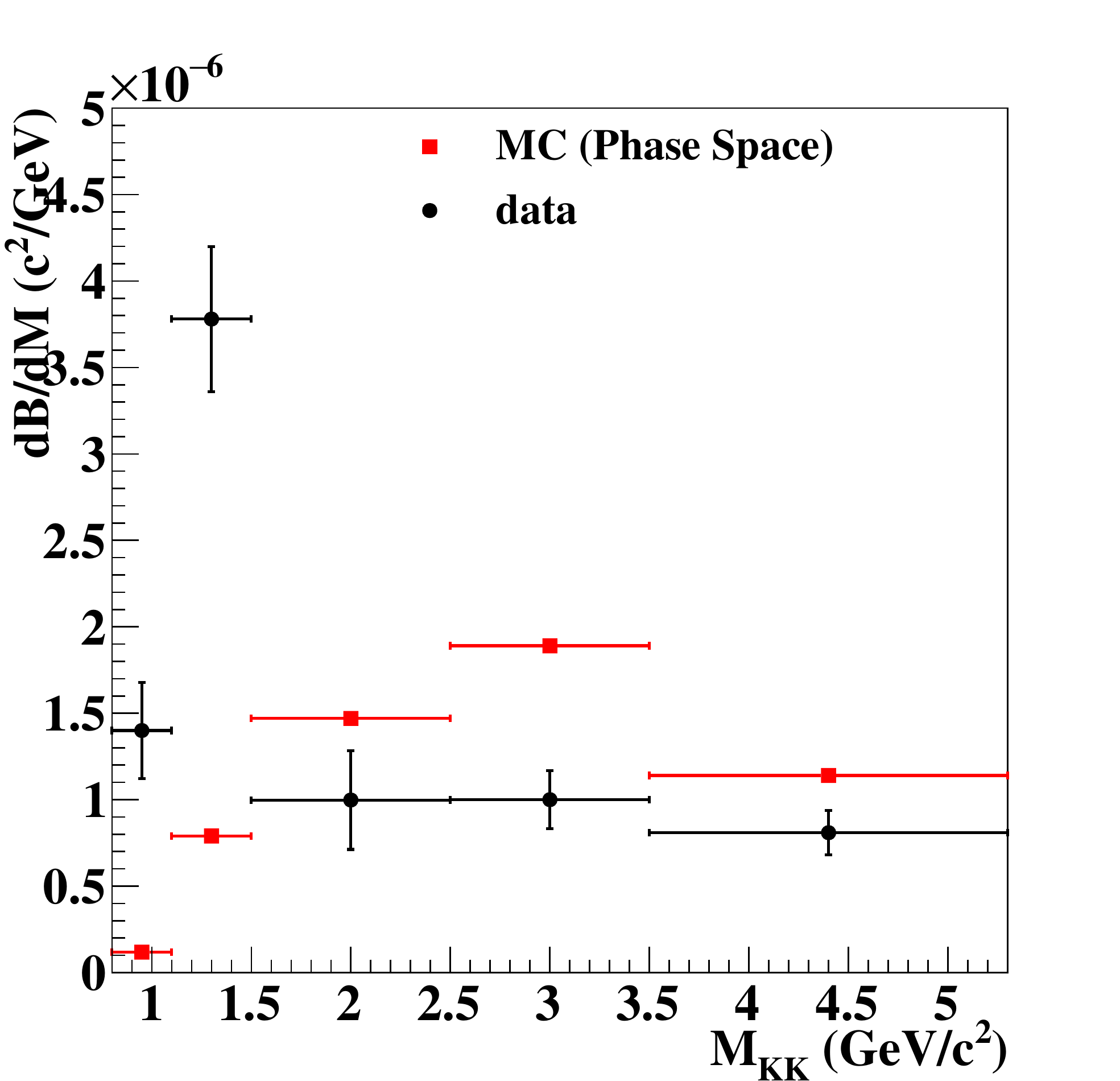}
\includegraphics[scale=0.25]{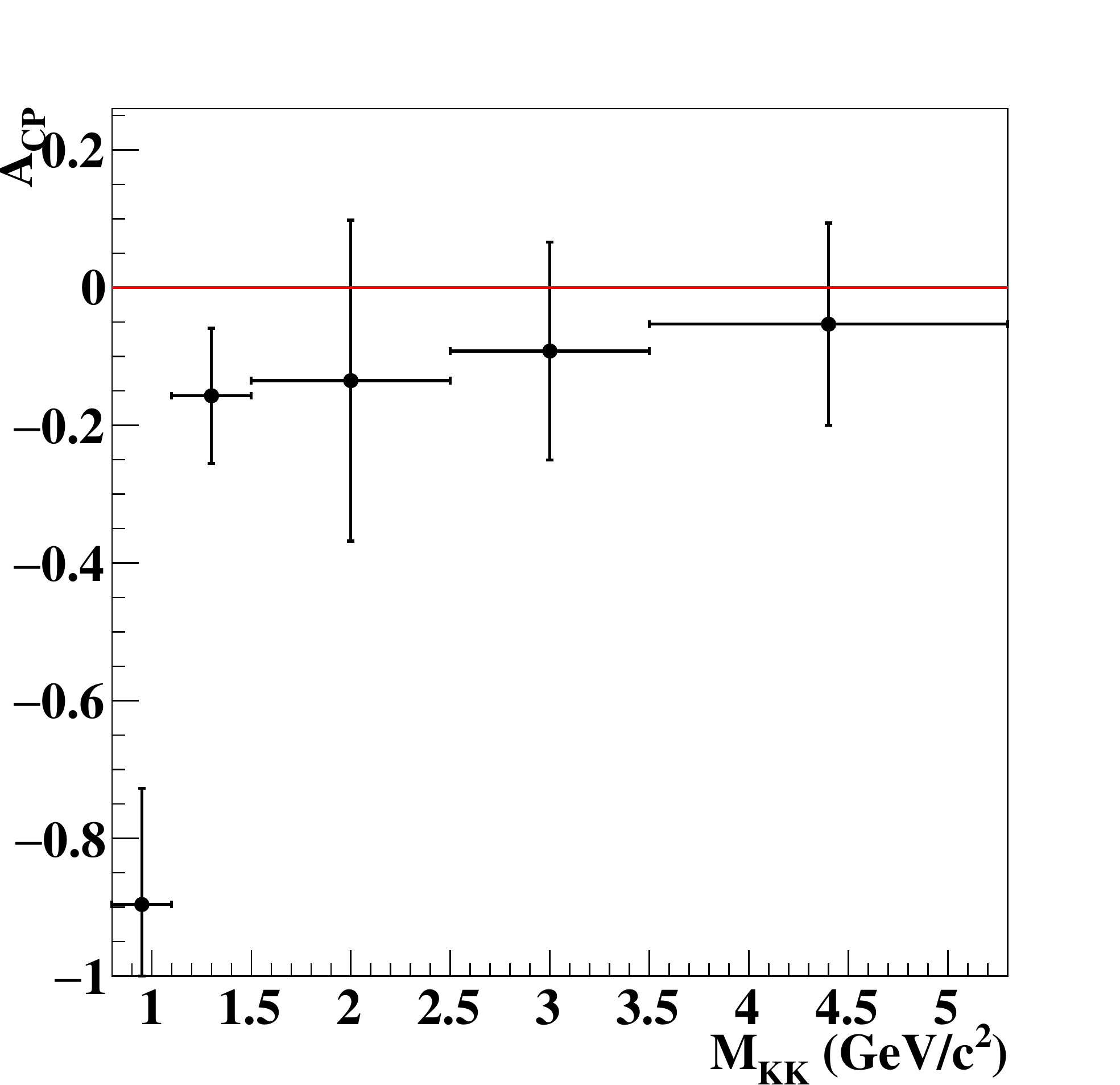}
\end{center}
\caption{Measured differential branching fractions (left) and $A_{CP}$ (right) as a function of $M_{K^{+} K^{-}}$. Each point is obtained from a two-dimensional fit with systematic uncertainty included. Red squares with error bars in the left plot show the expected signal distribution for a three-body phase space MC sample. Note that the phase space hypothesis is rescaled to the experimentally observed total $B^{+} \rightarrow K^{+} K^{-} \pi^{+}$ signal yield. 
}
\end{small}
\end{figure}

\section{$B^{0}\rightarrow \pi^{0}\pi^{0}$ decay}

One of the proposed techniques to measure $\phi_{2}$ is to perform an isospin analysis of the entire $B \rightarrow \pi \pi$ system [11]. This requires  measurements of $\mathcal{B}$ and $A_{CP}$  for $B^{+}\rightarrow \pi^{+}\pi^{0}$ and $B^{0}\rightarrow \pi^{0}\pi^{0}$ decays, along with that of $\mathcal{B}$ and time-dependent $CP$ asymmetry for the $B^{0}\rightarrow \pi^{+}\pi^{-}$ decay. One needs all these observables in order to  determine $\phi_{2}$ as electroweak tree and loop processes contribute with different phases to $B^{}\rightarrow \pi \pi$ decays. The $\mathcal{B}$ and $A_{CP}$ for $B^{0}\rightarrow \pi^{0}\pi^{0}$ are the least well determined among the $B^{}\rightarrow \pi^{}\pi^{}$ decays. This decay is also important to probe the disagreement between quantum-chromodynamics-based factorization, which predicts $\mathcal{B}$ below $1 \times 10^{-6}$ [12, 13], and previous measurements from Belle and BaBar of $(1.8 - 2.3) \times 10^{-6}$ [14,15]. Our study is based on a data sample recorded at the $\Upsilon(4S)$ resonance with the Belle detector comprising of $752 \times 10^{6}$ $B\bar{B}$ pairs, which corresponds to an integrated luminosity of 693 ~fb$^{-1}$, and an additional 83.35 ~fb$^{-1}$  recorded 60 MeV below the $\Upsilon (4S)$ resonance.

We reconstruct the signal $B^{0}$ candidate from a pair of $\pi^{0}$ candidates, each subsequently decaying to two photons. In addition to photons reconstructed from clusters in the electromagnetic calorimeter (ECL) that do not match any charged track, photons that convert to $e^{+}e^{-}$ pairs in the silicon vertex detector (SVD) are recovered and reconstructed as $\pi^{0} \rightarrow \gamma e^{+}e^{-}$. This provides a 5.3\% increase in detection efficiency. These photons must have an energy greater than 50 (100) MeV in the barrel (endcap) region of the ECL. The invariant mass of the two-photon combination
 must lie in the range $\mathrm{115}  < \mathrm{m_{\gamma \gamma} < 152 ~MeV/}c^{2}$, corresponding to $\pm 2.6 \sigma$ around the nominal $\pi^{0}$ mass [9]. As in the case of $B^{+} \rightarrow K^{+} K^{-} \pi^{+}$,  two kinematic variables $\Delta E$ and $M_{bc}$ are used to select the signal candidates. All candidates satisfying $M_{bc} > 5.26~ \mathrm{GeV}/c^{2}$ and $-0.3  < \Delta E < 0.2~\rm{GeV}$ are retained for further analysis. For 7.2\% of the events, there are multiple $B^{0}$ candidates in which case we choose the one that minimizes the deviation of the two $\pi^{0}$'s reconstructed invariant masses from the world average [9]. This criterion selects the true $B^{0}$ candidate in 90\% of MC events.

The dominant background is from  $e^+ e^- \rightarrow ~q\bar{q}~ (q= u, d, s, c)$ continuum process. To suppress this, we develop a Fisher discriminant ($T_{c}$) out of 16 modified Fox-Wolfram moments [16] combined  with the cosine of the polar angle of the $B$ candidate with respect to the z-axis and the cosine of the angle between the thrust axis of
the $B$ candidate and rest of event in the center-of-mass frame. All candidates with $T_{c}$ values below $-$0.3 are discarded, removing 72\% of the continuum background while retaining 98\% of signal events. Background events that arise from $b \rightarrow c$ transitions are mostly due to out-of-time ECL events originating from $e^+ e^-$ interactions, which leave large energy deposits in the ECL. This leads to a "pileup" event resembling a hadronic event with high energy back-to-back photons in the center-of-mass frame, and thus passes the first-level trigger. When combined with random photons from the hadronic interaction, they appear as two $\pi^{0}$'s with a large invariant mass. These events peak near the nominal $B$ mass [9] in $M_{bc}$. Since the events are recorded in coincidence with hadronic interactions, they also mimic $B$-like events in the continuum suppression variable $T_{c}$. A criterion on the trigger time of the ECL crystals, which selects ECL interactions in-time with the rest of the event, is employed to suppress this background removing 99\% of pileup at the cost of only 1\% of signal. The dominant  background from the rare $B$ decays ($b \rightarrow u, d, s$ transitions) is  due to $B^{+} \rightarrow \rho^{+} \pi^{0}$, where the charged pion from the subsequent $\rho^+ \rightarrow \pi^+ \pi^0$ decay is lost. This background peaks at similar values of $M_{bc}$ and $T_c$ as signal, but has $\Delta E$ shifted to negative values due to energy loss from the missing $\pi^+$.

The flavor of the reconstructed $B$ candidate is determined via a tagging procedure described in Ref. [17]. The tagging information is given by two parameters: the $b$-flavor charge q [+1 ($-$1) tagging a $B^{0}(\bar{B^{0}})$] and purity $r$. For the signal extraction, separate PDFs are constructed for the SVD1 (S1) and SVD2 (S2) data sets. We divide the data into seven bins, each for positive and negative tagged $r$-values, for both S1 and S2. The signal yield and $A_{CP}$ are extracted by performing a three-dimensional simultaneous unbinned extended maximum likelihood fit to the subsequent 28 data sets with $M_{bc}$, $\Delta E$ and $T_c$. Figure 3 shows the signal-enhanced projections of the fits to data in the three variables. We obtain a signal yield of $217 \pm 32$ events. The results are~[18]

\begin{equation}
\mathcal{B}(B^{0} \rightarrow \pi^{0} \pi^{0} ) = (1.31 \pm 0.19 \pm 0.18) \times 10^{-6} 
\end{equation} 
and
\begin{equation}
A_{CP} = +0.14 \pm 0.36 \pm 0.12, 
\end{equation} 
where the quoted uncertainties are statistical and systematic, respectively.

Combining our results for  $\mathcal{B}$ and $A_{CP}$ for $B^{0} \rightarrow \pi^{0} \pi^{0}$ with Belle's previous measurements of $\mathcal{B}$ and time-dependent $CP$ asymmetry for $B^{0}\rightarrow \pi^{+}\pi^{-}$  [19] and $\mathcal{B}$ and $A_{CP}$ for $B^{+} \rightarrow \pi^{+} \pi^{0}$  [20] allows us to employ the isopsin analysis of Ref. [11] to constrain $\phi_2$ . The result of the fit is shown in Fig. 4. Our results exclude $15.5^{\circ} < \phi_{2} <75.0^{\circ}$ at 95\% confidence level.
\begin{figure}
\begin{center}
\includegraphics[scale=0.45]{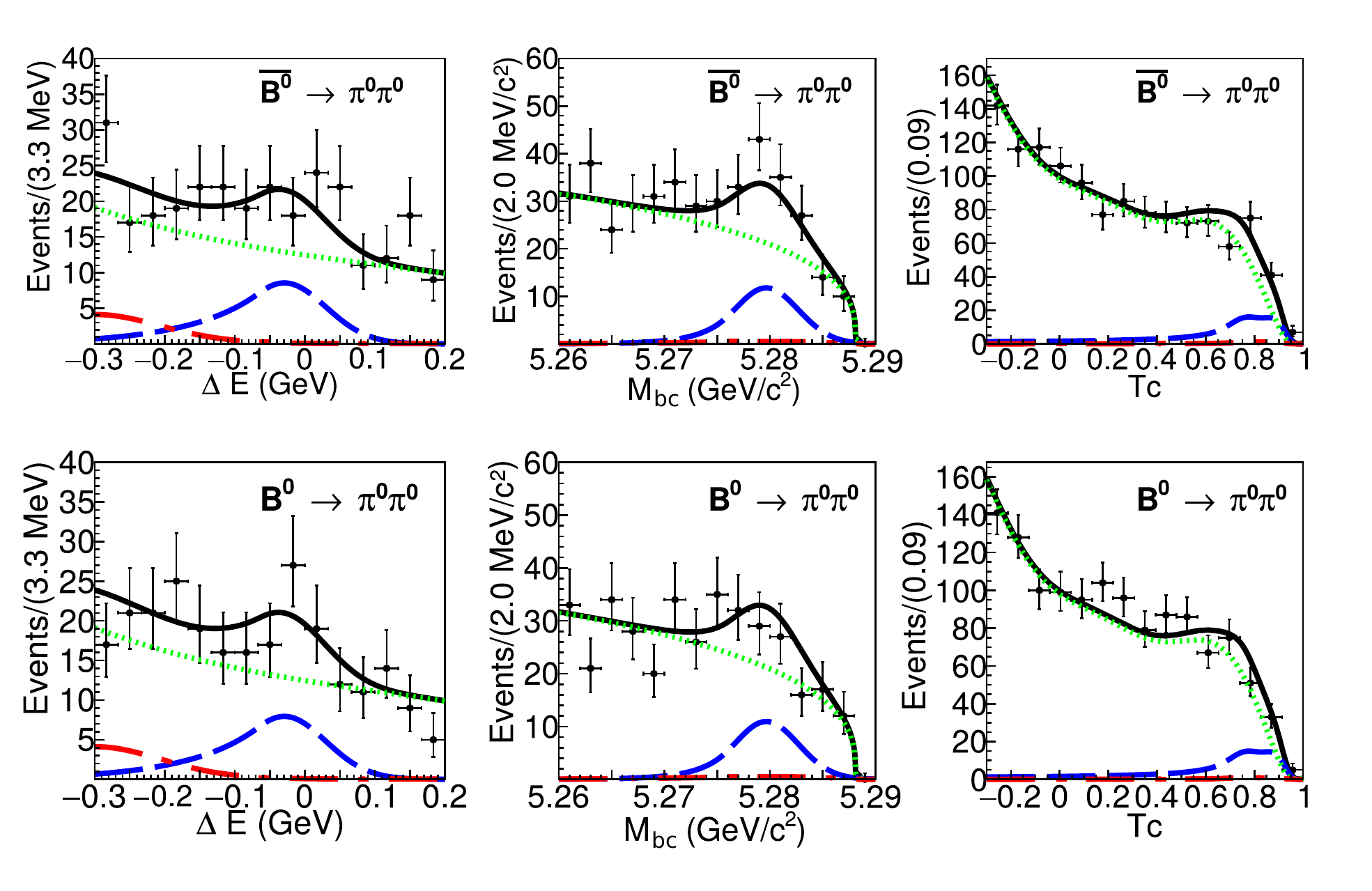}
\end{center}
\caption{Projections of the fit results onto (left) $\Delta E$, (middle) $M_{bc}$, and (right) $T_c$ . Data are points with error bars; the full fit results are shown by the solid black curves. Contributions from signal, continuum $q\bar{q}$, combined $\rho \pi$, and other rare $B$ decays are shown by the dashed blue, dotted green, and dash-dotted red curves, respectively. The top (bottom) row panels are for events with positive (negative) $q$ tags.}
\end{figure}
\begin{figure}[]
\begin{center}
\includegraphics[scale=0.4]{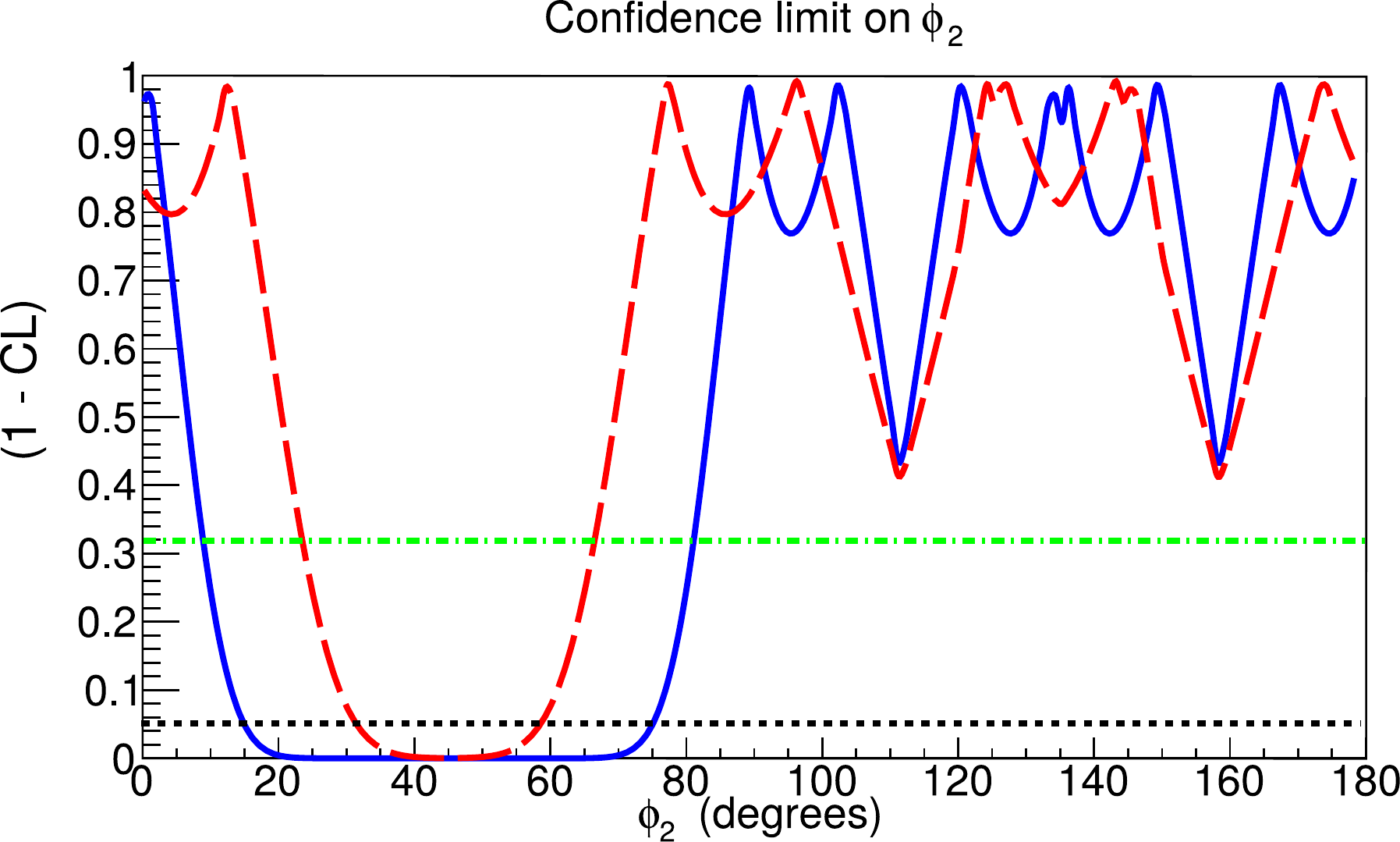}
\end{center}
\caption{Scan of the confidence level for $\phi_2$ using only data from $B^{} \rightarrow \pi^{} \pi^{}$ measurements of the Belle experiment. The dashed red curve shows the previous constraint from Belle data [19], while the solid blue curve includes our new results. The straight black dashed line is the 95\% confidence level and the green dot dashed line shows 68\% confidence level.}
\end{figure}\\

\section{Conclusion}
We have reported the recent measurements from Belle of the branching fractions and $CP$ asymmetries for $B^{\pm} \rightarrow K^{+} K^{-} \pi^{\pm}$ and $B^{0} \rightarrow \pi^{0} \pi^{0}$ decays using a data sample collected with the Belle detector. For the decay $B^{\pm} \rightarrow K^{+} K^{-} \pi^{\pm}$,  we confirm the observations by BaBar and LHCb, and find strong evidence for a large $CP$ asymmetry at the low $M_{K^+ K^-}$ region. Measurements of $\mathcal{B}$ and $A_{CP}$ in $B^{0} \rightarrow \pi^{0} \pi^{0}$  enable improved constraints on the angle $\phi_{2}$ of the CKM unitarity triangle.  Although the result is closer to theory predictions than the earlier Belle [14] and BaBar [15] measurements, it is still larger than expectations. The upcoming Belle II experiment [21], with its projected  50 times increased luminosity, will enable precision measurements of $\mathcal{B}$ and $A_{\rm{CP}}$ of $B^{0} \rightarrow \pi^{0} \pi^{0}$ and other $B^{} \rightarrow \pi^{} \pi^{}$ decays to strongly constrain $\phi_{2}$.

\section*{Acknowledgement}

We thank the KEKB group and all collaborating institutes and funding agencies that support the work of the members of
the Belle collaboration. We also extend our gratitude to Indian Institute of Technology Madras for the financial support to attend the conference.


\begin{thebibliography}{99}
 \bibitem{...} S. Kurokawa  and E. Kikutani, Nucl. Instr. Methods Phys. Res., Sect. A {\bf499}, 1 (2003); T.Abe et al., Prog. Theor.
Exp. Phys. {\bf2013}, 03A001 (2013).
\bibitem{...} Throughout this paper, the inclusion of the charge-conjugate decay modes is implied unless otherwise stated.
\bibitem{...} B. Aubert et al. (BaBar Collaboration), Phys. Rev. Lett. {\bf99}, 221801 (2007).
\bibitem{...} R. Aaij et al. (LHCb Collaboration), Phys. Rev. Lett. {\bf112}, 011801 (2014).
\bibitem{...} R. Aaij et al. (LHCb Collaboration), Phys. Rev. D {\bf90}, 112004 (2014).
\bibitem{...} B. Bhattacharya, M. Gronau, and J. L. Rosner, Phys. Lett. B {\bf726}, 337 (2013).
\bibitem{...} I. Bediaga, O. Lourenco, and T. Frederico, Phys. Rev. D {\bf89}, 094013 (2014).
\bibitem{...} M. Feindt and U. Kerzel, Nucl. Instr. Methods Phys. Res., Sect. A {\bf559}, 190 (2006).
\bibitem{...} C. Patrignani et al. (Particle Data Group), Chin. Phys.C {\bf40}, 100001 (2016).
\bibitem{...} C.-L. Hsu et al. (Belle Collaboration), Phys. Rev. D {\bf96}, 031101 (2017)
\bibitem{...} M. Gronau and D. London, Phys. Rev. Lett. {\bf65}, 33813384 (1990).
\bibitem{...} H. Li and S. Mishima, Phys. Rev. D {\bf73}, 114014 (2006).
\bibitem{...} H. Li and S. Mishima, Phys. Rev. D {\bf83}, 034023 (2011).
\bibitem{...} Y. Chao et al. (Belle Collaboration), Phys. Rev. Lett. {\bf94}, 181803 (2005).
\bibitem{...} J. Lees et al. (BaBar Collaboration), Phys. Rev. D {\bf87}, 052009 (2013).
\bibitem{...} G. C. Fox and S. Wolfram, Phys. Rev. Lett. {\bf41}, 1581 (1978); K. Abe et al. (Belle Collaboration), Phys. Rev. Lett. {\bf87}, 101801 (2001); K. Abe et al. (Belle Collaboration), Phys. Lett. B {\bf511}, 151 (2001).
\bibitem{...} H. Kakuno et al., Nucl. Instrum. Methods Phys. Res., Sect. A {\bf533}, 516 (2004).
\bibitem{...} T. Julius et al. (Belle Collaboration), Phys. Rev. D {\bf96}, 032007 (2017)
\bibitem{...} J. Dalseno et al. (Belle Collaboration), Phys. Rev. D {\bf88}, 092003 (2013).
\bibitem{...} Y.-T. Duh et al. (Belle Collaboration), Phys. Rev. D {\bf87},
031103(R) (2012).
\bibitem{...} T. Abe et al., arXiv:1011.0352 [physics.ins-det].

\end{thebibliography}
\end{document}